\input phyzzx
\nonstopmode
\sequentialequations
\tolerance=5000
\overfullrule=0pt
\nopubblock
\twelvepoint

\line{\hfill }
\line{\hfill IASSNS-HEP 96/28 }
\line{\hfill cond-mat/9604007}
\line{\hfill March 1996}

\titlepage

\title{Remarks on the Current-Carrying State of Hall
Superfluids\foot{Contribution to
the 1st Jagna International Workshop on Advances in Theoretical Physics.}}

\author{Frank Wilczek\foot{Research supported in part by DOE grant
DE-FG02-90ER40542.~~~wilczek@sns.ias.edu}}

\vskip.2cm
\centerline{{\it School of Natural Sciences}}
\centerline{{\it Institute for Advanced Study}}
\centerline{{\it Olden Lane}}
\centerline{{\it Princeton, N.J. 08540}}
\endpage

\abstract{I discuss in an elementary and self-contained way the nature
of the current-carrying state in Hall superfluids.  The
explicit connection of fractional charge and rationality  of the
filling fraction $\nu$
with quantization of angular momentum,
and the discussion of the preferred velocity profile for bulk current
flows 
may
be new; the latter might provide an interesting experimental handle on a
fundamental aspect of quantum superfluidity.}

\endpage

\REF\girvin{R. Prange and S. Girvin, ed. {\it The Quantum Hall
Effect\/} (Springer-Verlag, 1987).
The chapters by Girvin are particularly closely
related to the material discussed here, and are highly recommended.}

\REF\stone{M. Stone, ed. {\it Quantum Hall Effect\/} (World
Scientific, 1992).}

The quantum Hall effect exhibits many analogies to the more venerable
superfluidities, epitomized by the $\lambda$ phase of $^4$He and BCS
superconductivity.  The most obvious macroscopic commonality
is of course the possibility of (approximately)
dissipationless flow.  At a more microscopic and theoretical level,
there are striking analogies between the Laughlin quasiparticles and
vortices, and effective Landau-Ginzburg like theories using
interesting concepts including fictitious gauge fields and
Chern-Simons interactions which
conveniently summarize important aspects of the phenomenology
[\girvin , \stone ].  

Here I
would like to focus on one particular question which lies at the heart
of every form of superfluidity, and can be discussed in a
self-contained and (I hope) instructive fashion: that is, the nature
of the ideal current-carrying state and  particularly its realization in a
non-simply connected geometry, where fundamental questions of
quantization come into play.  I have not seen a discussion of quite 
this
type for the Hall superfluid, but that may well just reflect ignorance
on my part.

\chapter{Approach Through Wave Functions}

A remarkable feature of the theory of the quantum Hall effect is how
much insight can be gained by consideration of simple trial wave
functions.  I will review, partially explain, and exploit this feature
in the following.

\section{Quick March Through Droplets and Annuli}

Let me remind you of some basic results about electrons in a strong
magnetic field.  Here and throughout I will assume that the motion of
the electrons is confined to a plane.  I will also ignore the spin
degree of freedom.  The energy levels are highly
degenerate Landau levels, with a density of states $(2\pi l_B^2 )^{-1}$ per unit
area per Landau level, where the magnetic length $l$ is defined
through
$l_B^{-2} \equiv eB/\hbar c$.  The splitting between levels is $\hbar$
times the cyclotron frequency, {\it viz}. $\Delta E = \hbar (eB/mc)$.
At low temperatures and for densities small compared to $(2\pi l_B^2)^{-1}$ it
ought to be a good approximation to restrict attention to states
formed from single-particle states taken from the lowest Landau level,
unless there is some very special energetic advantage to admixing
higher levels.  Within the lowest Landau level, the single particle
wave functions take a particularly attractive form if one employs the
so-called symmetric gauge, defined by the vector potentials $A_x =
By/2, A_y = -Bx/2$.  With this gauge choice, the wave functions in the
lowest Landau level take the form
$$
\psi = f(z)e^{-{1\over 4}|z|^2}~,
\eqn\lll
$$
where $f(z)$ is an arbitrary holomorphic ({\it i.e}. complex analytic)
function of $z\equiv x+ iy$,
subject to reasonable growth conditions insuring 
that the wave function is
normalizable.  Distances are measured in units of the magnetic length. 
A basis of orthogonal vectors in this Hilbert space is provided by the
functions
$f_l(z) = z^l$.  $l$ is the canonical angular momentum around the
origin, which here is intrinsically non-negative.  For reasonably
large $l$, the corresponding wave function is concentrated in a
circular ring of radius $\sqrt {2l}$ and width $\sqrt {2\pi} $ centered
at the origin.  It follows, by comparing the size of the region where
the wavefunction is large to the  inverse density, or by direct
calculation, that the supports of these wave functions are highly
overlapping.

Now let us consider an assembly of (non-interacting) electrons.  Let
us suppose that they are subject to a very small potential which draws
them toward the origin, but does not appreciably change the form of
the wave functions.  Then the ground state will be formed by occupying 
the states with the smallest values of $l$, consistent with Fermi
statistics.
It will be the Slater determinant
$$
\psi_1 = \det \{ z_r^{c-1} \} e^{-{1\over 4} \sum |z_k|^2 }~,
\eqn\wfnct
$$
where the row variable $r$ (a particle identity index),  the column
variable $c$, and $k$ all run from 1 to $N$, the number of electrons.
Given the spatial character of the wave functions as discussed above,
one easily sees that $\psi_1$ represents, for large values of $N$, a
droplet of uniform density $1/2\pi$ and radius $\sqrt {2N}$, with some
fuzziness in an ring of width unity near the edge.  The uniformity of
the density is far from obvious upon inspecting \wfnct , but
follows from the fact that we have occupied all the lowest Landau
level states with substantial support 
in the specified geometric region, since the completely occupied
Landau level is certainly uniform.  The deep reason why
this uniformity is not 
more obvious is that to maintain any fixed gauge condition, such as
the symmetric gauge we have chosen,  
a translation must be accompanied by a non-trivial gauge
transformation.   
One can also
re-write the droplet wave function 
in a suggestive way using the Vandermonde identity
$\det \{ z_r^{c-1} \} = \prod_{k<l; k,l = 1}^N (z_k - z_l)$.    

Part of Laughlin's inspiration in understanding the fractional
quantized Hall effect was to notice that the cube of this wave
function has remarkable qualities, that make it a particularly
attractive trial wave function for an assembly of interacting
electrons.  After taking the cube the Gaussian factor is no longer
appropriate to the lowest Landau level, but that defect can be
compensated by a trivial redefinition of the length scale, which we
suppose done.  Then clearly one has a wave function again describing a
uniform droplet centered at the origin, now with radius $\sqrt
{6N}$, density $1/6\pi$ (that is, filling factor $1/3$) and
fuzziness in an ring of width $\sqrt {6\pi }$ at the edge.  From the
alternative Vandermonde representation of the determinantal wave
function as a product, we see that there is a strong vanishing as two
particles approach one another.  Thus we might expect -- and one does
find, through numerical work -- that this is a particularly favorable wave
function for short-range repulsive interactions among the electrons.

Now consider, in either the integer or fractional case, the process of
adiabatically turning on a localized point source of flux.  (Flux comes in
tubes in three space dimensions, and therefore
as localized points in two space
dimensions.)  Let us implement the potential as a branch cut.
In this situation the equations which forced  
$f(z)$ to be holomorphic 
(in the
lowest
Landau level)
still hold in the bulk, but a singularity is allowed along the cut.
For adiabatic evolution, since there is a gap, one will not leave the
lowest Landau level, and the wave functions will be multiplied by
factors $z^\alpha$, with $\alpha$ proportional to the flux.  
The effect of 
adiabatically adding a quantum unit $h/e$ 
of flux at the origin is to multiply the
single particle wave functions by a factor $z$, 
an operation which maps us between proper lowest
Landau level wave functions.  From our previous discussion of the
one-particle wave functions, we see that the effect of this factor 
is to push
them out from the origin.  The effect on the many-particle 
droplet wave function is to produce a localized 
hole in the charge distribution.  
Proceeding similarly and adding many units of flux, we will
carve out a large geometrical hole in the charge droplet, producing an
annulus of uniform charge density.

Thus far I have been speaking of free electrons.  It 
would impede my narrative flow to give an 
exposition of the fundamentals of the quantum Hall effect, 
but I
would like to say just enough to make it plausible why the sort of ``wave
functionology''
discussed here has some relevance for describing the real phenomena.
The two most profound features of the observed phenomena are the
existence of incompressible states of the electron fluid in a strong
magnetic field, and the fact that these incompressible states 
occur when the ratio of the density to the density required to fill
the Landau level is a simple rational fraction.  For example, the
most robust and prominent such states occur at 
$\nu \equiv \rho 2\pi l_B^2 = 1, 1/3$.  The key physical point for
incompressibility is the absence of low-energy excitations -- ``sound
waves'' --
characterized
by  small
deviations from the favored density over a large spatial region.  
The rigidity of the ground-state wave function against density
variations arise  because correlations fix a preferred 
length scale.  In general this is
not a single-particle phenomenon.  
That is clear for $\nu =1/3$:
in the absence of interactions a one-third filled Landau level
would remain hugely degenerate, since one could pick and choose among
many states to occupy, 
and in this case is easy to construct ``sound wave''
configurations.
Exactly at $\nu=1$ all the available states are filled, and there is
indeed a finite energy 
gap to density fluctuations, since these require promotion of
some electron to the second Landau level.  However this argument fails at
any infinitesimal distance from $\nu=1$, and does not fully explain the
observed phenomenon, which cannot and does not require infinite
precision.  The full explanation requires careful consideration of
impurities, and is well beyond my scope here (see [\girvin ]).  

Despite these essential complications the simple (Laughlin) 
wave functions we have been discussing are profoundly correct and relevant.
The reason lies in the nature of the quantized Hall state.
Since the ground state is characterized by an energy gap, a set of 
trial wave
function with the correct short-range correlations and a sensible
thermodynamic limit can have
significant overlap with the true ground state and capture its
universal features.  There is ample numerical work indicating that the
Laughlin wave functions, with their built-in feature of uniformity and
strong short-range repulsion,
make an excellent approximation to the true
ground states.  

Likewise adiabatic flux insertion, which as we have seen creates a
localized hole in the charge distribution, produced a good wave
function for the low-energy charged excitations -- quasiholes.
Adiabatically inserting many flux 
units at the origin produces a macroscopic hole, and provides
a good trial wave
function for annular geometry.

\chapter{Adiabatic Charge Insertion and Angular Momentum}

Now I would like to reconsider in the context of the Hall effect the
analysis of flow in an annular geometry, which plays a classic role in
the discussion of quantization for the other superfluidities.  I will
do this first in the style of the preceding section, using explicit
wave functions.

\section{Adiabatic Charge Insertion}

Let us now consider adiabatic charge insertion -- or, more
precisely, response to an adiabatic change in the electric potential.
This proves to be subtly but crucially different
from adiabatic flux insertion.  We can consider adiabatic charge
insertion as arising from the slow lowering of charge down toward the
sample plane.  For simplicity let us consider lowering a negative
charge toward the origin.

One's first reaction might be that the droplet or annulus responds to
such a perturbation by carving out a bigger hole, as we discussed for
flux insertion.  This is not quite adequate, however.
During  the charge lowering  process the degree of the wave function, 
regarded as a
polynomial in the angular variables $e^{i\theta_j}$, cannot change.
Indeed, since this degree is necessarily an integer it can only vary
adiabatically if it does not change at all.  This defect can be
remedied in a simple manner by applying together with the old factor
$$
{\rm Annulizing ~ Factor~} = \prod z_k^p
\eqn\annulizing
$$
a new factor
$$
{\rm Rotation ~ Factor~} = \prod e^{-ip\theta_k}~.
\eqn\rotation
$$
At the moment this probably appears to be nothing but an {\it ad hoc\/}
fix; but I hope to convince you there is more to it.

\section{Angular Momentum}

\REF\feynman{R. Feynman, R. Leighton, and M. Sands {\it The Feynman
Lectures on Physics\/} (Addison-Wesley, 1963) vol. II, Secs. 17-4 and 27-6.}

Before proceeding further 
let me remind you of some very elementary but sometimes 
confusing facts about angular
momentum, which justify the terminology above and will be of 
use below.  The canonical angular momentum is associated with the
generator of spatial rotations or the operator
$-i{\partial \over \partial \phi}$; its value is $l$ on a wave function
whose angular dependence is $e^{il\phi}$.  The canonical angular momentum of
a particle is conserved in its response to azimuthally symmetric
external fields.  The kinetic angular momentum operator is defined
with covariant derivatives, and is gauge invariant.  It is what
appears in energy expressions, is what gravitons couple to, ... and in
general is what we think of as a proper measure of rate of rotation.
The sum of the
kinetic angular momentum of particles and the field angular momentum of the
electromagnetic field is conserved.
In symmetric gauge in the lowest Landau level all the wave functions
$z^l e^{-{1\over 4}|z|^2}$ have unit kinetic angular momentum,
corresponding to cyclotron motion in a small orbit, independent of
$l$.  The factor \rotation\ implements a dynamical change in the
kinetic angular momentum, by a fixed amount $\hbar p$ for each particle.

It may appear paradoxical that a procedure (lowering the charge source)
which can be carried out in
a completely rotation-invariant way could cause rotational motion of the
electron fluid.  However this ``paradox'' is closely related to a classic
one in electromagnetic theory [\feynman ] and it is resolved in a similar 
way, very instructive for our purpose.  Radial expansion of the
annulus
involves transport of charge and thus 
a change in the electric field and eventually a change in the field
angular momentum, as we shall now compute. 

For a constant magnetic field, ${\bf B}= B {\bf\hat z}$
and an electric potential, $V$, with azimuthal symmetry,
the total field angular
momentum is in the $z$ direction and 
has magnitude
$$ 
L_{\rm EM} = {1 \over 4 \pi c} 2 \pi B \int dz r d r \;
r {\partial V \over \partial r} . 
\eqn\Lemcylind
$$
Notice that this diverges for a point charge at the origin.
 
Consider then a point charge $+Q$ at the origin and a ring of
charge $-Q'$ at radius $r_0$. 
Using the
expansion of  $1/|{\bf x} - {\bf x}'|$
appropriate to cylindrical coordinates we obtain 
$$
\eqalign{ L_{\rm EM} = &- {2 BQ \over c} \int_0^{\infty} dk \; k \delta
(k)  
\int_0^\infty d r r^2 I_0 (0) K_0'(k r) \cr 
 &  - {2 BQ' \over c} \int_0^{\infty} dk \; k \delta
(k)  \left\{
\int_0^{r_0} d r r^2 I_0' (k r) K_0 (k r_0) +
\int_{r_0}^\infty d r r^2 I_0 (k r_0) K_0'(k r) 
\right\} , \cr}  
\eqn\Lchargeplusring
$$
where $f' (x)$ denotes ${df \over dx}$. 
Doing the $r$ integrals by parts and making use of the
expansions
$I_0 (x) = 1 + {x^2 \over 4} + O(x^4), K_0 (x)  = -
\ln \alpha x (1 + {x^2 \over 4} ) + {x^2 \over 4} + O(x^4 \ln x)$
(where $\alpha=0.5772\ldots$) 
to evaluate the resulting $k$--integrand for small $k$, we find
$$
L_{\rm EM} = - {2 BQ \over c} \int_0^{\infty} dk \;  \delta
(k) \left\{ {2 (Q-Q') \over k^2} + {Q' r_0^2 \over 2} \right\}
. 
$$ 
When $Q=Q'$ the divergent pieces at $k=0$ cancel 
and we find the
simple result 
$$
L_{\rm EM} = - {B Q r_0^2 \over 2c } . 
\eqn\Lfinal 
$$ 
It is important that this arises as 
the finite residual of an infrared divergence,
since this indicates the insensitivity of the result to details, including
notably the 
precise shape of the edge. 

Of course, if we move the annulus so as to move charge $Q$ from the
inner radius $r_1$ to the outer radius $r_1$,
the change in electromagnetic field angular momentum is 
${BQ (r_0^2 - r_1^2)\over 2c}$.  Since the number of particles in the
annulus is given by $N= 2\pi (r_0^2 - r_1^2 ) \rho $ with 
$\rho =  {\nu e B\over 2\pi \hbar c}$, we find the relation 
$$
{\Delta L \over N} ~=~ {\hbar Q \over e\nu }
\eqn\partamom
$$
for the angular momentum per particle.  

Note that according to \partamom\ a unit change in 
angular momentum corresponds to a generally fractional change 
$e\nu $ in the ring charge $Q$.  Thus we see how fractional charge can
be transported from one edge of the sample to the other.  Also note
that for transport of an integer number of electrons across the ring
to be possible, $\nu$ must be rational.  Evidently  elementary
considerations
on angular momentum and its quantization 
are tied up with the most profound quantum
aspects of the Hall states.


\chapter{Analogies and Identification}

In this section I will argue, both
by analogy and by  energetics, that one should
identify the product of the standard annulus wave function with a
rotation factor of type \rotation\ as the
preferred current-carrying Hall state.

\section{The London-Feynman-Girvin Description}

The Hall fluid at rest is 
supposed to be described by a unique featureless ground state wave
function $\Phi (x_k)$, that has a finite energy gap against localized
density inhomogeneities.  Overall translational motion without change
of density is 
then described by  wave
functions of the type
$$
\Psi_q (x_k) ~=~ \prod e^{iqx_k} \Phi (x_k)~.
\eqn\qwf
$$
One may also consider configurations 
in which $q$ becomes a slowly
varying function of position, $q\rightarrow q(x)$.  These again do not
violate the incompressibility constraint, and ought to represent bulk
motions of the fluid.  Following Feynman's approach to superfluid
helium, one can work to extract dynamics directly from these
wave-functions.

An alternative approach involves the concept of a macroscopic wave
function or order parameter, as pioneered by London and by Landau and
Ginzburg.
In this approach one postulates the existence of a macroscopic complex
wave
function $\Upsilon (x)$ which describes the ordered state and its
low-energy excitations.   
One expresses the dynamics directly in terms of $\Upsilon (x)$.  The
ground state is characterized by a uniform constant value of
$\Upsilon$, and the energy functional for the low-energy excitations
is supposed to be expanded in gradients, 
${\cal H} \approx \kappa |\partial \Upsilon |^2$.  In London's work the
magnitude of $\Upsilon$ was taken constant, and this will be good
enough for my immediate purpose here. 

Girvin and his collaborators adapted this approach  to the Hall fluids
in a very fruitful way [\girvin ], and here I am just filling out a
particular corner of their picture.

We have two alternative descriptions of the current-carrying states, 
which is potentially one too many.  But it is easy to see the relation: with 
$\Upsilon (x ) = {\rm const.~}e^{iq(x)}$ we are describing \qwf , and 
{\it vice versa}.  To firm up the connection we should calculate the
energy associated with \qwf\ and check that it takes the assumed form,
in the process determining  $\kappa$.

With this perspective we come to view \rotation\ in a different way.
If there is an underlying London theory of the
Hall superfluids, then this must be the way to describe the
ideal current-carrying state for annular geometry.   
To check this identification, let us consider the energetics.  Having
appended a factor $r^p$ to each single-particle wave function
(combining the annular and rotation factors), the resulting
single-particle state is no longer within the lowest Landau level.
However for large $l$ one can compute the relevant overlap to be
$$
{\int_0^\infty r^{2l+p} e^{-r^2/2} r dr \over
\sqrt {\int_0^\infty r^{2l} e^{-r^2/2} r dr 
  \int_0^\infty r^{2l+2p} e^{-r^2/2} r dr }} \approx 1- {p^2\over 8l}
\eqn\overlap
$$
The energy will thus be of order the cyclotron energy times the sum of
$p^2 \over 4l$ taken over the relevant $l$-values for the annulus,
that is for between $l_i \approx r_i^2 $ at the inner radius and 
$l_o \approx r_o^2$ at the outer.  We see that this sum goes
quadratically with the overall current strength and logarithmically
with the radius.  But this precisely matches what we expect from the
macroscopic wave function approach.   Indeed the current density 
is of the
form
$\Upsilon^* ({-i\over r} {\partial \over \partial \theta }) \Upsilon
\sim p/r$ and
the energy density is proportional to the square of this, with 
$\Upsilon \sim e^{ip\theta }$.

\section{Profile}

\REF\lw{R. Levien and F. Wilczek, paper in preparation.}

\REF\levien{R. Levien, C. Nayak, and F. Wilczek {\it Space-Time
Aspects of Quasiparticle Propagation\/} Int. J. of Mod. Phys. {\bf B9}
3189 (1995).}

From this analysis we conclude that the ideal current
density will exhibit a $1/r$ profile.  Indeed, this is an immediate consequence
of the from of the rotation factor \rotation .  From the point of view
of the macroscopic wave function, it is a signature of condensation in
a definite partial wave.  From the single-particle point of view, it
expresses the conservation of circulation.  

There is an absolutely essential aspect of the problem which has not
been
mentioned so far.  That is, when I discussed the bulk translational
motion 
and
considered the effect on the wave function \qwf , I failed to take
into account that the whole process is going on in a background
magnetic field.   Because of this background, a proper implementation
of the boost will require the existence of an appropriate electric
field.  We really only reproduce a configuration related locally by
boosts to the ground state when the 
famous Hall relation 
$j/E ~=~ \nu e^2/h$ 
between the current and perpendicular electric field is obeyed.   

If the
ideal current profile we discussed above is achieved, then the Hall
relation gives a simple definite result for $E(r)$ -- a
conveniently measured quantity.  We were led to infer its form by assuming
that the bulk flow occurred with uniform density, and (implicitly)
that the voltage took care of itself.  
Neither of these assumptions is automatically fulfilled. 
The
presence of Laughlin quasiholes, which cost only a finite energy, will
generally both spoil the density uniformity and help in reaching a
self-consistent 
potential.  However one should note that the $j(r) \propto 1/r$ profile
besides maintaining strict uniformity in
correlated flow also minimizes the kinetic energy
for a given total current.  This is more favorable than the situation
in superfluid helium, where the natural constraint is given total
angular momentum.  With that constraint, the most favorable profile is
instead $j(r) = {\rm const.}$

We are actively investigating the question 
under what practical conditions, if any, the ideal flow is a good
approximation [\lw ].

Acknowledgment:  A brief 
indication of these ideas occurs at the
end of [\levien ].  The talk as actually given was closely based on
that paper; here I have given a more self-contained and
extended discussion of one aspect.
Anything original in this talk represents joint work with 
R. Levien and C. Nayak, from whom I have learned much regarding these matters.

\endpage

\refout

\end